# Extreme Value Statistics in Silicon Photonics

David Borlaug and Bahram Jalali

Electrical Engineering Department, UCLA, Los Angeles, CA 90095

*Abstract*— **We show that fluctuations of Raman amplified pulses, in the presence of a noisy pump, follow extreme value statistics, and provide mathematical insight into the origin of this perplexing behavior.**

L-shape probability distributions are extremely non-Gaussian functions that have been surprisingly successful in describing the frequency of occurrence of extreme events ranging from stock market crashes and natural disasters, the structure of biological systems and fractals, and optical rogue waves [1]. They are an example of extreme value statistics, a class of stochastic distributions in which events much larger than the mean (outliers) can occur with significant probability. In stark contrast, the ubiquitous Gaussian distribution heavily favors events close to the mean, and all but forbids highly unusual and cataclysmic events.

In this paper we report the observation of extreme value statistics in silicon photonics. We experimentally show that the distribution of Raman amplified pulses, in the presence of a noisy pump, follows power law statistics. We find that 16% of the Stokes pulses account for 84% of the pump energy transfer, an uncanny resemblance to the empirical Pareto principle or the 80/20 rule that describes important observation in socioeconomics. We also describe a model that provides physical and mathematical insights into such unusual but fascinating behavior.

The experiments were performed at mid-IR wavelengths on a 2.5 cm thick bulk silicon sample. The pump laser was a Q-switched Nd:YAG pumped OPO emitting pulses at a wavelength of 2.9 microns. The pulse width as reported by the manufacturer is approximately 4 ns (FWHM), but it is known that the envelope is highly asymmetric and has pronounced intra-pulse structure. The Stokes was a CW HeNe laser at 3.4 microns. After traveling through the sample, the beams are separated using a dichroic beam splitter. Further isolation between the two beams is achieved using a spectrometer. The pump wavelength was detected with a photodetector (Coherent J25LP-MB) with a response time of 80 microseconds. The Stokes wavelength was detected with an InAs photodetector with a response time of 15 ns. The detectors were not fast enough to resolve the intra-pulse structure of the envelope; however, this is of no concern because we are merely interested in pulse energy fluctuations. Measuring these fluctuations only requires pulse-to-pulse relative energy measurements, for which the detectors are adequate. The detected electrical signal is then recorded on an oscilloscope which is triggered at the pump repetition rate.

Figure 1 shows the histogram of 3000 pulses for the pump pulse energy. As can be seen the distribution is nearly symmetric around a well defined mean energy – in other words, it has the general features of an ordinary (Gaussian or Rician) distribution.

The Raman gain experienced by the Stokes signal is obtained by comparing the output Stokes peak pulse energy in the presence and absence of the pump. Again, only relative pulse-to-pulse statistical fluctuations are of interest here. The histogram (3000 pulses) of the Raman gain is shown in Figure 2. Since the Stokes input is the same for all pulses, the gain distribution also describes the distribution of amplified Stokes pulses. The observed distribution clearly shows L-shape extreme value behavior, highlighted by the high probability of large outliers in the "fat-tail" of the distribution. While most pulses experience modest gain, a small percentage of pulses experience amplification far larger than that of average events. This is even more evident in the log scale plot shown in the inset. The distribution shows that the number of such pulses does not decay exponentially with their amplitude, as would be expected for a Gaussian distribution, but rather has a power-law dependence, as expected for an extreme value random process. While the number of extreme value pulses is small, their contribution to the total energy cannot be ignored.

It has previously been shown that the Rician distribution [2] is the proper model for amplitude fluctuations arising from a coherent field perturbed by narrowband noise [3-4]. We invoke it here as an approximation to fluctuations of the pump laser. Mathematically, the Rician distribution is given by,

$$f(E_p) = \frac{E_p}{\sigma^2} Exp\left[-\frac{E_p^2 + \upsilon^2}{2\sigma^2}\right] I_0\left[\frac{E_p \upsilon}{\sigma^2}\right] \quad (1)$$

where $E_p$ is the electric field *amplitude* for the pump laser, $I_0$ is the modified Bessel function of the first kind with order zero, $\upsilon$ and $\sigma^2$ are parameters that influence the offset and width of the distribution. The function approaches a Rayleigh distribution in the limit when $\upsilon \to 0$, evident in Figure 3. Through a nonlinear transformation, the corresponding pump intensity distribution is given by,

$$f(I_p) = \frac{Z_0}{4n\sigma^2} Exp\left[-\frac{Z_0 I_p / 2n + \upsilon^2}{2\sigma^2}\right] I_0\left[\frac{\upsilon}{\sigma^2}\sqrt{\frac{Z_0 I_p}{2n}}\right] \quad (2)$$

Here, $I_p = 2nE_p^2 / Z_0$, where $Z_0$ is the impedance of free space and $n$ is the refractive index of the medium,

To show how extreme value behavior emerges from the seemingly innocuous Rician distribution we start with the simple expression for the Raman gain,

$$G = e^{g \cdot I_p \cdot L} \quad (3)$$

here $g_R$ is the Raman gain coefficient and L is the pump-Stokes interaction length. This nonlinear function causes the pump intensity fluctuations, given by Equation 2, to be transformed into the following distribution for the Raman gain,

$$f(G) = G^{-(1+b/2\sigma^2)} \cdot \left(\frac{b}{2\sigma^2} Exp\left[-\frac{\upsilon^2}{2\sigma^2}\right]\right) \cdot I_0\left[\frac{\upsilon\sqrt{bLn(G)}}{\sigma^2}\right]$$

(4)

where $b \equiv Z_0/2ng_R L$. The first term in this expression readily predicts that the Raman gain will have a power law behavior, the hallmark of extreme value statistics. This is graphically shown in Figure 4.

The inability to measure the pulse envelope and its intra-pulse structure limits the measurement to that of energy fluctuations, whereas, the model is based on the intensity fluctuations. This prevents the quantitative comparison between the experiments and the model. However, the model does unambiguously explain the salient feature of the experimental observations, and shows how extreme value statistics, for the Stokes, emerges from transformation of the pump intensity statistics by the nonlinear relation between gain and pump intensity. Although a Rician distribution was chosen to represent the pump fluctuations, this choice is not critical. Extreme value behavior for the Stokes emerges from Gaussian pump distribution as well, as long as pump is noisy enough.

In summary, we have shown that the fluctuations of Raman amplified Stokes signal in the presence of a noisy pump follow extreme value statistics and have provided mathematical insight into its origin. Our observations have important practical implications. For example, a few Stokes pulses carry most of the converted beam energy. Also, the extreme deviation from Gaussian statistics implies that the traditional characterization based on the standard deviation will provide a highly incomplete description of the pulse to pulse fluctuations.

Acknowledgments: This work was supported by Dr. Henryk Temkin of DARPA-MTO. The authors thank Drs. D. Dimitropoulos and D.R. Solli of UCLA for discussions.

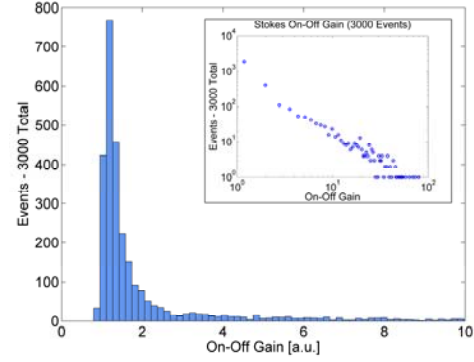

Fig. 2: Measured histogram of amplified Stokes pulses. Inset shows the same data plotted on the log scale. Extreme value statistical behavior and outlier events are clearly evident in the tail of the L-shape distribution.

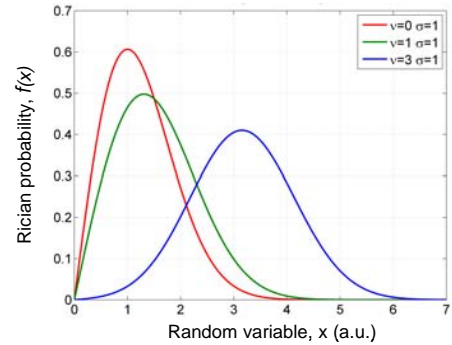

Fig. 3: The Rician distribution.

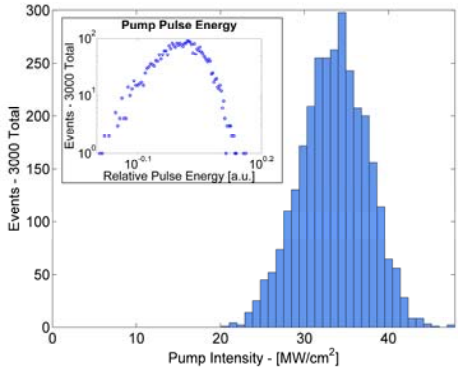

Fig. 1: Measured histogram of pump pulses. Inset shows the same data plotted on the log scale.

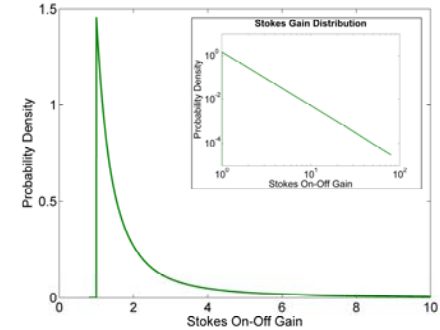

Fig. 4: Distribution of amplified Stokes pulses according to the model. Inset shows the same plotted on the log scale. The model produces the salient features of experimental data shown in Fig. 2.